\documentclass[11pt,a4paper]{article}

\usepackage{amsmath,amssymb}
\usepackage{a4wide}

\newcommand{\BE}{\begin{equation}}
\newcommand{\EE}{\end{equation}}
\newcommand{\cE}{\mathcal{E}}
\newcommand{\cF}{\mathcal{F}}
\newcommand{\cI}{\mathcal{I}}
\newcommand{\cO}{\mathcal{O}}
\newcommand{\bM}{\mathbb{M}}
\newcommand{\mR}{\mathbb{R}}
\newcommand{\e}{\mathrm{e}}
\newcommand{\BB}{\mathbf{B}}
\newcommand{\ff}{\mathbf{f}}
\newcommand{\hh}{\mathbf{h}}
\newcommand{\ee}{\mathbf{e}}

\newcommand{\KK}{\mathbf{K}}
\newcommand{\xx}{\mathbf{x}}
\newcommand{\qq}{\mathbf{q}}
\newcommand{\yy}{\mathbf{y}}
\newcommand{\pp}{\mathbf{p}}
\newcommand{\uu}{\mathbf{u}}
\newcommand{\vv}{\mathbf{v}}
\newcommand{\nnu}{\mathbf{\nu}}
\newcommand{\ssg}{\sigma}
\newcommand{\zzero}{\mathbf{0}}
\newcommand{\bk}[1]{{\langle #1 \rangle}}
\newcommand{\abs}[1]{{\vert {#1} \vert}}
\newcommand{\pt}{\partial}
\newcommand{\eq}{\mathit{me}}
\DeclareMathOperator{\BF}{\cI}
\DeclareMathOperator{\TR}{Tr}

\begin{document}
\title{{\huge Hydrodynamic equations \\ for an electron gas in graphene}}
\author{L. Barletti \\
{\small Dipartimento di Matematica e Informatica ``Ulisse Dini''} \\
{\small Viale Morgagni 67/A, 50134 Firenze, Italy}\\
\tt{ \small luigi.barletti@unifi.it}
}
\date{}
\maketitle
\abstract{In this paper we review, and extend to the non-isothermal case,
some results concerning the application of  
the maximum entropy closure technique to the derivation of  hydrodynamic
equations for particles with spin-orbit interaction and Fermi-Dirac statistics.
In the second part of the paper we treat in more details the case of electrons on a graphene 
sheet and investigate various asymptotic regimes.}
\section{Introduction}
\label{S1}
This paper is devoted to present some results on the derivation of hydrodynamic equations
describing electrons subject to spin-orbit-like interactions.
Systems of this kind, of particular interest for applications to microelectronics,  
include electrons undergoing the so-called Rashba effect \cite{Zutic02}, the 
Kane's two-band K$\cdot$P model \cite{Kane66} and electrons in single-layer graphene 
\cite{CastroNeto09}.
The diffusive and hydrodynamic descriptions of such systems are extensively treated in 
Refs.\ \cite{JMP10,JSP10,PossannerNegulescu11,JMP14,MorandiBarletti14,BarFroMor14,BarBorFro14}.
Here, we summarize the results contained in Refs.\ \cite{JMP14,MorandiBarletti14,BarFroMor14,BarBorFro14},
 concerning the hydrodynamic description, and extend them to the non-isothermal case.
\par
In comparison with kinetic models, the advantages of fluid models for applications are evident. 
In fact, from the numerical point of view, a system of PDEs for a set of macroscopic quantities is much more desirable
than a single equation for a density in phase-space, where also the components
of momentum are independent variables. 
Moreover, from the point of view of mathematical modeling, they offer more flexibility, as various 
kind of boundary conditions and coupling terms (e.g. with a self-consistent
potential or with various types of scattering mechanisms) can be very naturally embodied in the
mathematical model.
\par
On the other hand, the derivation of fluid equations for the systems under
consideration, which possess spinorial degrees of freedom and non-parabolic dispersion relations
(energy bands), is far from being a trivial extension of the techniques employed for standard 
(i.e.\ scalar and parabolic) particles.
The best strategy to obtain hydrodynamic (or, more in general, fluid-dynamic) equations in this case, 
seems to be their systematic derivation from an underlying kinetic description by means of the
Maximum Entropy Principle (MEP) and its quantum extensions 
\cite{Levermore96,DR03,TrovatoReggiani10,CR14}.
The MEP basically stipulates that the microscopic (kinetic) state of the system is the most probable among all
states sharing the same macroscopic moments of interest, providing therefore a formal closure of the
system of moment equations.
This is a very general principle which finds a variety of applications to different fields, 
ranging from statistical mechanics to signal theory \cite{Wu97}. 
For quantum systems it can be used in combination with the quantum kinetic framework provided by the
phase-space formulation of quantum mechanics due to Wigner \cite{ZachosEtAl05,Bumi03}.
\par
In the present work the Wigner formalism is used ``semiclassically'', which means that some
quantum features are retained (namely, the peculiar energy-band dispersion relations and the Fermi-Dirac statistics)
while others are neglected (namely, the quantum coherence between different bands).
Consequently, the obtained hydrodynamic description misses some interesting physics  
when quantum interference between 
bands becomes important (e.g.\ close to abrupt potential variations \cite{KatsnelsonEtAl06,CheianovEtAl2007}).
Nevertheless, the derived equations possess an interesting mathematical structure and 
reveal some interesting physics, still occurring in absence of such interference phenomena (see, in particular,
Section \ref{S5}). 
\par 
The paper is organized as follows.
In Section \ref{S2} the kinetic-level formalism, based on a semiclassical Wigner description, is introduced for 
a fairly general spin-orbit Hamiltonian that includes all cases of interest. 
In Sec.\ \ref{S3} we write the moment equations for density, velocity and energy, and perform 
their formal closure by means of the MEP.
Then, the second part of the paper is focused on the  case of graphene.
In Sec.\ \ref{S4} the general theory exposed in the first part is specialized for the Dirac-like
Hamiltonian describing electrons on a single-layer graphene sheet. 
In Sec.\ \ref{S5} we obtain the asymptotic form of the hydrodynamic equations derived in Sec.\ \ref{S4} 
in some physically relevant limits (namely, the high temperature, zero temperature, collimation and 
diffusive limits).
Finally,  Sec.\ \ref{S6} is devoted to conclusions and perspectives.
\section{Phase-space description of spin-orbit particles}
\label{S2}
Let us consider a spin-orbit Hamiltonian of the form
\BE
\label{H}
 H(\xx,\pp) = \left[h_0(\pp) + V(\xx) \right]\sigma_0 + \hh(\pp) \cdot \ssg,
\EE
where $\xx \in \mR^d$, $\pp \in \mR^d$, $\ssg = (\sigma_1,\sigma_2,\sigma_3)$, $\hh = (h_1,h_2,h_3)$ 
and
$$
 \sigma_0 = \begin{pmatrix} 1 & 0 \\ 0 & 1   \end{pmatrix},
\quad
 \sigma_1 = \begin{pmatrix} 0 & 1 \\ 1 & 0   \end{pmatrix},
\quad
 \sigma_2 = \begin{pmatrix} 0 &-i \\ i & 0   \end{pmatrix},
\quad
  \sigma_3 =  \begin{pmatrix} 1 & 0 \\ 0 &-1   \end{pmatrix}.
$$
Moreover, the dot product is defined as $\hh \cdot \ssg = h_1\sigma_1 + h_2\sigma_2 +h_3\sigma_3$.
\par
Hamiltonians of this kind describe various systems of great interest in solid-state physics.
The first example is a 2-dimensional electron gas confined in an asymmetric potential well, which 
is subject to the Rashba spin-orbit interaction \cite{Zutic02,JMP10}. 
In this case:
$$
d = 2,
\qquad
h_0(\pp) = \frac{1}{2m^*} \abs{\pp}^2,
\qquad
\hh(\pp) = \alpha \pp \times \ee_z,
$$
where $\pp = (p_x,p_y)$, $m^*$ is the electron effective mass, $\ee_z$ is the normal direction to the well 
and $\alpha$ is the Rashba constant.
\par  
Another example is the two-band K$\cdot$P model \cite{Kane66,BarFroMor14,BarBorFro14,BarNBA11};
in this case:
$$
d = 3,
\qquad
h_0(\pp) = \frac{1}{2m} \abs{\pp}^2,
\qquad
\hh(\pp) = \left(0,\, \frac{\hbar}{m}\KK\cdot\pp,\, \frac{E_g}{2}  \right), 
$$
where $m$ is the electron (bare) mass, $E_g$ is the band-gap and $\KK$ is the matrix element 
of the gradient operator between conduction and valence Bloch functions.
\par
The last example is that of electrons on a single-layer graphene sheet 
\cite{CastroNeto09,JMP14,DeretzisLaMagna14}, in which case:
$$
d = 2,
\qquad
h_0(\pp) = 0,
\qquad
\hh(\pp) = c\pp.
$$
This case will be considered in more details in the second part of the paper.
\par
\smallskip
We remark that the variable $\pp$ has to be interpreted as the crystal pseudo-momentum, 
rather than the ordinary momentum.
The interpretation of the vector variable $\ssg$ depends on the cases: it is (proportional to) 
the spin vector in the case of Rashba 
Hamiltonian, while it is a pseudo-spin in the other two examples \cite{Kane66,CastroNeto09}.
For graphene, in particular, the pseudo-spin is related to the decomposition of the honeycomb lattice
into two inequivalent sublattices, which reflects the presence of two carbon atoms in the fundamental
cell of the lattice \cite{CastroNeto09,Weiss58}.
\par
The main semiclassical quantities associated with \eqref{H} are:
\begin{enumerate}
\item
the two energy bands 
\BE
  E_\pm(\pp) = h_0(\pp) \pm \abs{\hh(\pp)}
\EE
(i.e., the eigenvalues of $H$ with $V=0$);
\item 
the projectors on the eigenspaces corresponding to $E_\pm(\pp)$,
\BE
P_\pm(\pp) = \frac{1}{2} (\sigma_0 \pm \nnu(\pp)\cdot\ssg),
\EE
where 
\BE
\nnu(\pp) = \frac{\hh(\pp)}{\abs{\hh(\pp)}}
\EE
is the pseudo-momentum direction;
\item
the semiclassical velocities
\BE
\label{vel}
  \vv_\pm (\pp) = \nabla_{\pp} E_\pm(\pp);
\EE
\item
the effective-mass tensor \cite{BarNBA11}
\BE
\label{emass}
 \bM_\pm^{-1}(\pp) = \nabla_{\pp}\otimes \vv_\pm (\pp)
 = \nabla_{\pp}\otimes\nabla_{\pp} E_\pm(\pp).
\EE
\end{enumerate}
Note, in particular, that the eigenvalues of the projector $P_\pm$ are 1 and 0, corresponding  to whether or not  
the electron energy belongs to the upper/lower energy band.
Hence, the expected value of $P_\pm$  can be interpreted as the fraction of electrons belonging to 
the upper/lower band (see below).
\par
\smallskip
The phase-space description of a statistical population of electrons with Hamiltonian \eqref{H} is provided
by the Wigner matrix \cite{BarFroMor14,ZachosEtAl05,Bumi03,Morandi09}
\BE
\label{Fdef}
F(\xx,\pp,t) = \sum_{k=0}^3 f_k(\xx,\pp,t) \sigma_k,
\EE
which is the Wigner transform,
$$
  f_k(\xx,\pp,t) = \int_{\mR^d} 
  \rho_k\left( \xx + \frac{\qq}{2},  \xx - \frac{\qq}{2} \right) \e^{-i\qq \cdot \pp/\hbar}  d\qq,
$$
of the spinorial density matrix
$$
  \rho(\xx,\yy,t) = \sum_{k=0}^3 \rho_k(\xx,\yy,t) \sigma_k.
$$
Such representation of a quantum mixed-state has the fundamental property that the expected value
of an observable  with symbol 
$A = \sum_{k=0}^3 a_k(\xx,\pp) \sigma_k$ is given by the classical-looking formula
\BE
\label{ExVal}
\mathbb{E}_F[A] = \int_{\mR^{2d}}  \TR(FA) d\xx\,d\pp = \frac{2}{(2\pi\hbar)^d} \int_{\mR^{2d}}  \sum_{k=0}^3 a_k(\xx,\pp)\, f_k(\xx,\pp,t)\, d\xx\,d\pp.
\EE
By applying Eq.\ \eqref{ExVal} to the band projectors $P_\pm(\pp)$ we obtain 
$$
\mathbb{E}_F[P_\pm] = \frac{1}{(2\pi\hbar)^d} \int_{\mR^{2d}} (f_0 \pm \nnu\cdot\ff)\,d\xx\,d\pp 
$$
and it is therefore natural to interpret the functions 
\BE
\label{fpm}
  f_\pm = f_0 \pm \nnu\cdot\ff
\EE
as the phase-space densities of electrons having energies, respectively, in the upper and lower band.
\par
Let us now consider the following hydrodynamic moments of electrons in the two bands:
\BE
\label{moments}
\begin{aligned}
  &n_\pm = \bk{f_\pm},& \quad &\text{(density),}
\\[6pt]
  &n_\pm\uu_\pm = \bk{\vv_\pm \,f_\pm}& \quad &\text{(velocity),}
\\[6pt]  
  &n_\pm e_\pm = \bk{E_\pm\,f_\pm}& \quad &\text{(energy),}
  \end{aligned}
\EE
(see also Ref.\ \cite{CR14} where additional moments are considered).
Here we have introduced the shorthand
$$
\bk{f}(\xx,t)  = \frac{1}{(2\pi\hbar)^d} \int_{\mR^d} f(\xx,\pp,t)\,d\pp.
$$
The (semiclassical) dynamics of the Wigner matrix \eqref{Fdef} is provided by the 
Wigner equations for the Hamiltonian \eqref{H} \cite{BarFroMor14},
\BE
\label{WE}
\left\{
\begin{aligned}
 &\left( \pt_t   + \nabla_\pp h_0\cdot\nabla_\xx - \nabla_\xx V \cdot \nabla_\pp \right) f_0
+ \sum_{k=1}^3 \nabla_\pp h_k\cdot\nabla_\xx f_k= 0,
\\[4pt]
 &\left( \pt_t   + \nabla_\pp h_0\cdot\nabla_\xx - \nabla_\xx V \cdot \nabla_\pp \right) f_i
 +\nabla_\pp h_i \cdot\nabla_\xx f_0
 =  \frac{2}{\hbar}\, (\hh \times \ff)_i, 
\end{aligned}
\right.
\EE
with $i = 1,2,3$.
From \eqref{WE}, the following equations for the band-Wigner functions $f_+$ and $f_-$ (see
definition \eqref{fpm}) are readily obtained:
\BE
\label{WE2}
  \left( \pt_t  + \vv_\pm \cdot\nabla_{\xx}  -\nabla_{\xx} V\cdot\nabla_{\pp} \right) f_\pm  = 
  - \nabla_{\xx}\cdot \ff_\perp
  \pm \nnu\cdot(\nabla_{\xx} V\cdot\nabla_{\pp})\ff_\perp,
\EE
where the terms containing 
$$ 
  \ff_\perp :=  (\nnu \times \ff)\times \nnu
$$ 
are responsible for quantum interference between the two bands \cite{BarFroMor14,Morandi09}.
\section{Maximum entropy closure}
\label{S3}
In order to obtain from \eqref{WE2} a closed system of equations for the moments \eqref{moments},
we assume that the system is in a state $F^\eq$ of maximum entropy,
according to the so-called {\em Maximum Entropy Principle (MEP)} \cite{Levermore96,DR03,TrovatoReggiani10,CR14,Wu97} 
which in the present case reads as follows:
\par
\medskip
\noindent
{\bf MEP\ }
{\em $F^\eq$ is the most probable microscopic state with the observed 
macroscopic moments $n_\pm$, $\uu_\pm$ and $e_\pm$.}
\par 
\medskip
\noindent
Hence, we search for a Wigner matrix $F^\eq$ that maximizes the total entropy
\BE
\label{TE}
  \cE(F) = -\frac{k_B}{(2\pi\hbar)^d} \int_{\mR^{2d}} \TR\{ s(F)\}(\xx,\pp)\, d\pp\,d\xx
\EE
among all matrices $F = \sum_{k=0}^{3} f_k\sigma_k$,  such that $0 \leq F \leq 1$ and
\BE
\label{constraints}
  \bk{ \begin{pmatrix} 1 \\ \vv_\pm \\ E_\pm \end{pmatrix} f_\pm} = 
  n_\pm \begin{pmatrix} 1 \\ \uu_\pm \\ e_\pm \end{pmatrix}.
\EE
In \eqref{TE}, $k_B$ is the Boltzmann constant, $\TR$ is the matrix trace and 
\BE
\label{entro}
  s(x) =  x\log x + (1 - x)\log(1- x), \qquad 0 \leq x \leq 1,
\EE
is (minus) the Fermi-Dirac entropy function.
The condition $0 \leq F \leq 1$ ensures that $s(F)$ is a well-defined matrix.
\par
It can be proven \cite{JSP10} that 
\begin{multline}
\label{MEPform}
  f_\pm^\eq = (s')^{-1} \left(\vv_\pm \cdot\BB_\pm  + A_\pm - C_\pm E_\pm\right)
\\[3pt]
  = \frac{1}{\exp\left(C_\pm E_\pm - \vv_\pm \cdot\BB_\pm  - A_\pm\right)+1},
\end{multline}
where $A_\pm$, $\BB_\pm = (B_1,\ldots,B_d)_\pm$ and $C_\pm$ 
are Lagrange multipliers (functions of $\xx$ and $t$),
and, moreover,
\BE
\label{NoInt}
  \ff_\perp^\eq = \zzero.
\EE
Thus, the (semiclassical) MEP state corresponds to two local Fermi-Dirac 
distributions in the two energy bands.
In particular, Eq.\ \eqref{NoInt} implies that, in such state, the interference terms 
vanish and, therefore, the two bands are decoupled (unless additional coupling mechanisms
are considered \cite{PossannerNegulescu11,BarBorFro14,CR14}).
Hence, from now on, we shall treat the two bands separately and, in order to simplify notations, the
$\pm$ labels will be suppressed (except where a distinction between quantities taking different forms 
in the two bands, such as $E_\pm$ or $\vv_\pm$, is necessary).
\par
Using \eqref{MEPform} and \eqref{NoInt} in \eqref{WE2} (and suppressing the $\pm$ labels, 
as it was just explained) yields
\BE
\label{WSC}
  \left( \pt_t  + \vv_\pm \cdot\nabla_{\xx}  -\nabla_{\xx} V\cdot\nabla_{\pp} \right) f^\eq  = 0.
\EE
By taking the moments $\bk{\cdot}$, $\bk{\vv_\pm \cdot}$ and  $\bk{E_\pm \cdot}$ of both sides of 
Eq.\ \eqref{WSC}, and recalling the definitions \eqref{vel} and \eqref{emass},  we obtain
the moment equations
\BE
\label{ME}
\left\{
\begin{aligned}
&\pt_t n  + \pt_j (n u_j) = 0,
\\[3pt]
&\pt_t (n u_i) + \pt_j P_{ij}^\pm   +  Q_{ij}^\pm \pt_j V = 0,
\\[3pt]
&\pt_t (n e) + \pt_j S_j^\pm + nu_j \pt_jV  = 0
\end{aligned}
\right.
\EE
where $\pt_i = \pt/\pt x_i$ and
\BE
\begin{aligned}
\label{PQ}
 &P^\pm_{ij} = \bk{v^\pm_iv^\pm_j f^\eq},  
 \\[4pt]
 &Q^\pm_{ij} = \bk{\textstyle{\frac{\pt v^\pm_i }{\pt p_j}} \,f^\eq} =  \bk{ (\bM_\pm^{-1})_{ij} \,f^\eq},
 \\[4pt]
 &S_j^\pm =  \bk{ E_\pm v_j^\pm \,f^\eq}.
\end{aligned}
\EE
Thanks to the MEP, the moment system \eqref{ME} is implicitly closed by the constraints
\eqref{constraints}, linking the Lagrange multipliers $(A,\BB,C)$ to the moments $(n,\uu,e)$
thus allowing (in principle) to think to $f^\eq$ as being parametrized by $(n,\uu,e)$ and,
consequently, the extra moments $(P^\pm,Q^\pm,S^\pm)$ as
functions of the unknowns $(n,\uu,e)$.
\par
Following Levermore \cite{Levermore96}, we can express the moments of the MEP state $f^\eq$
as the derivatives with respect to the Lagrange multipliers of the  ``density potential'' $\varepsilon^*$,
which is defined as the Legendre transform of the entropy density
$$
   \varepsilon = \bk{s(f^\eq)}
$$
where $s$ is given by \eqref{entro} and $f^\eq$ by  \eqref{MEPform}.
It is not difficult to show that
$$
  \varepsilon^* = -\bk{\log(1 - f^\eq)}
$$
and that the constraint equations \eqref{constraints} may be rewritten as
\BE
\frac{\pt \varepsilon^*}{\pt A} = n,
\qquad
  \frac{\pt \varepsilon^*}{\pt B_i} = nu_i,
\qquad
 -\frac{\pt \varepsilon^*}{\pt C} = ne,
\EE
where  $i = 1,\ldots,d$.
Levermore's theory, moreover, ensures that system \eqref{ME}, with the closure relations \eqref{PQ}
is hyperbolic and, therefore, it is at least locally well-posed (see also Ref.\ \cite{JMP14}).
\section{The case of graphene}
\label{S4}
We now specialize the formalism introduced so far to the case of a population of electrons on
a single-layer graphene sheet.
Such electrons, in the proximity of a Dirac point in pseudo-momentum space \cite{CastroNeto09},  
are described by the Hamiltonian \eqref{H} with
$$
d = 2,
\qquad
h_0(\pp) = 0,
\qquad
\hh(\pp) = c\pp,
$$
(where $c \approx 10^6\mathrm{m}/\mathrm{s}$ is the Fermi velocity), 
which corresponds to a Dirac-like Hamiltonian for relativistic, massless particles.
We remark that this is an approximation which is valid only in the proximity of a Dirac point 
for an infinite, ideal and un-doped system (see Ref.\ \cite{DeretzisLaMagna14} and references
therein).
In this case the energy bands are the Dirac cones
\BE
\label{cones}
E_\pm(\pp) = \pm c\abs{\pp},
\EE
and the eigenprojections are given by
\BE
P_\pm(\pp) = \frac{1}{2} (\sigma_0 \pm \nnu(\pp)\cdot\ssg),
\EE
where
\BE
\label{nudef2}
\nnu(\pp) = \frac{\pp}{\abs{\pp}}.
\EE
Moreover, the semiclassical velocities are
\BE
 \vv_\pm(\pp) =  \pm \frac{c \pp}{\abs{\pp}}  \pm c\,\nnu(\pp),
\EE
implying that electrons travel with the constant speed $c$ and direction $\nnu$, 
and the effective-mass tensor is 
\BE
\bM_\pm^{-1}(\pp) =  \frac{c}{\abs{\pp}} \, \nnu_\perp(\pp) \otimes \nnu_\perp(\pp)
\EE
where 
$$
\nnu_\perp = (-\nu_2,\nu_1).
$$
Since the lower band is unbounded from below, we have to change a little the theory developed 
in the previous sections and describe the lower-band population in terms of electron vacancies, i.e.\ {\em holes}.
This is achieved by means of the substitution 
$$
   f_-(\xx,\pp,t) \; { \longmapsto} \; 1 - f_- (\xx,-\pp,t),
$$
which brings the transport equation, Eq.\ \eqref{WSC}, into
\BE
\label{MEPgraph}
  \left( \pt_t  + c \,\nnu \cdot\nabla_{\xx}  \mp \nabla_{\xx} V\cdot\nabla_{\pp} \right) f^\eq  = 0.
\EE
Note that the only difference between electrons and holes is the charge sign.
Moreover, the MEP-states for electrons and holes have now the form
\BE
\label{MEPstate}
  f^\eq = \frac{1}{\exp\left(C\abs{\pp} - \nnu(\pp) \cdot\BB  - A \right)+1},
\EE
in fact, both upper-cone electrons and  lower-cone holes have positive energies 
\BE
\label{conic}
E(\pp) = c\abs{\pp}
\EE
(note that in \eqref{MEPgraph} the Fermi velocity $c$ has been absorbed in the Lagrange
multiplier $C$).
Moreover, we  slightly change the definition of $\uu$ to be the average {\em direction} 
$$
  n\uu = \bk{\nnu f}, \qquad \quad 0 \leq \abs{\uu} \leq 1,
$$
which differs from average velocity just for the constant factor $c$.
The inequality $\abs{\uu} \leq 1$ is an obvious consequence of the fact that $\uu$
is an average of directions.
\par
The moment equations \eqref{ME}, in the specific case of graphene, read as follows:
\BE
\label{MEgraph}
\left\{
\begin{aligned}
&\pt_t n  + c\pt_j (n u_j) = 0,
\\[3pt]
&\pt_t (n u_i) + c\pt_j P_{ij}   \pm  Q_{ij} \pt_j V = 0,
\\[3pt]
&\pt_t (n e) + c\pt_j S_j \pm c nu_j \pt_jV  = 0,
\end{aligned}
\right.
\EE
where the higher-order moments $P_{ij}$, $Q_{ij}$ and $S_j$ take the form
\BE
\label{PQR}
\begin{aligned}
&P_{ij} =  \bk{\nu_i\nu_j f^\eq},
\\[3pt]
&Q_{ij} = \bk{ \frac{1}{\abs{\pp}}\nu_i^\perp\nu_j^\perp f^\eq},
\\[3pt]
&S_j = \bk{cp_jf^\eq}.
\end{aligned}
\EE
\par
We now intend to find an (as much as possible) explicit expression for the dependence of the Lagrange multipliers 
$A$, $\BB = (B_1,B_2)$ and $C$ in terms of the moments $n$, $\uu = (u_1,u_2)$ and $e$, as resulting from the constraint equations
\BE
\label{CE}
  \bk{f^\eq} = n, \qquad \bk{\nnu f^\eq} = n\uu, \qquad \bk{c \abs{\pp} f^\eq} = ne.
\EE
By expressing the integrals over $\pp \in \mR^2$ in polar coordinates,
we obtain the expressions
\BE
\begin{aligned}
 &\bk{f^\eq} = \frac{\BF_0^2(A,\abs{\BB})}{2\pi\hbar^2 C^2},
\\[4pt]
 & \bk{\nnu f^\eq} = \frac{\BF_1^2(A,\abs{\BB})}{2\pi\hbar^2 C^2} \, \frac{\BB}{\abs{\BB}},
\\[4pt]
 &\bk{c \abs{\pp} f^\eq} =  \frac{c\BF_0^3(A,\abs{\BB})}{\pi\hbar^2 C^3},
\end{aligned}
\EE
where
\BE
\label{Idef}
  \BF_N^s(x,y) = \frac{1}{\pi}  \int_0^{\pi}  \cos(N\theta)\,\phi_s(x+y\cos\theta)\,d\theta,
\EE
and $\phi_s$ is the Fermi integral of order $s>0$:
$$
\phi_s(z) = \frac{1}{\Gamma(s)} \int_0^\infty \frac{t^{s-1}}{e^{t-z} + 1}\,dt.
$$
It is now convenient to put
\BE
\label{nTdef}
B = \abs{\BB}, \qquad C = \frac{c}{k_BT}, \qquad n_T = \frac{k_B^2 T^2}{2\pi\hbar^2c^2} = \frac{1}{2\pi\hbar^2 C^2},
\EE
so that the previous expressions can be rewritten as 
\BE
\label{exprlast}
\begin{aligned}
 &\bk{f^\eq} =n_T \BF_0^2(A,B)
\\[4pt]
 & \bk{\nnu f^\eq} = \frac{n_T}{B} \BF_1^2(A,B)\,\BB,
\\[4pt]
 &\bk{c \abs{\pp} f^\eq} =  2n_Tk_B T \BF_0^3(A,B).
\end{aligned}
\EE
We remark that the new Lagrange multiplier $T$ has the physical meaning of the electron gas temperature.
From \eqref{exprlast} and the constraint equations \eqref{CE}, we obtain that $\BB$  has the same direction as $\uu$ and
that $(n,\abs{\uu},e)$ are related to the scalar Lagrange multipliers $(A,B,T)$ by
\BE
\label{CE2}
\begin{aligned}
  &\BF_0^2(A,B)n_T = n, 
\\[4pt]
  &\frac{\BF_1^2(A,B)}{ \BF_0^2(A,B)} = \abs{\uu},
\\[4pt] 
  &\frac{\BF_0^3(A,B)}{ \BF_0^2(A,B)}2k_BT = e.
\end{aligned}
\EE
Similarly to what is found in Ref.\  \cite{JMP14}, we obtain the following expressions of the higher-order moments 
\eqref{PQR} in terms of $n$, $\uu$, $T$ and the functions $\BF_N^s=\BF_N^s(A,B)$:
\BE
\label{PQexpr}
\begin{aligned}
 &P_{ij} = \frac{n}{\abs{\uu}^2}\big( \frac{\BF_0^2+\BF_2^2}{2\BF_0^2} \, u_i u_j  
 + \frac{\BF_0^2-\BF_2^2}{2\BF_0^2} \, u_i^\perp u_j^\perp \big),
\\[4pt]
 &Q_{ij} = \frac{c}{k_BT}\frac{n}{\abs{\uu}^2}\big(\frac{\BF_0^1-\BF_2^1}{2\BF_0^2}\, u_i u_j 
 + \frac{\BF_0^1+\BF_2^1}{2\BF_0^2}\, u_i^\perp u_j^\perp \big),
\\[4pt]
&S_j = \frac{2k_BTn}{\abs{\uu}}\,\frac{\BF_1^3}{\BF_0^2} \,u_j,
\end{aligned}
\EE
where $\uu^\perp = (-u_2,u_1)$.
\section{Asymptotic regimes}
\label{S5}
The expressions \eqref{PQexpr} of $P_{ij}$, $Q_{ij}$  and $S_j$ 
are still not explicit, as functions of  $n$ and $\uu$.
In fact, these expression depend, through the functions $\BF_N^s(A,B)$,
on the two scalar Lagrange multipliers $A$ and $B$, which are related to $n$ and $\abs{\uu}$
via the relations \eqref{CE2}.
In Ref.\ \cite{JMP14} it has been proven that the correspondence between $(A,B)$ and $(n,\abs{\uu})$
is 1-1 but, as far as we know, it is not possible to give an explicit, analytic, expression of the former as 
functions of the latter.
\par
However, we can say more in some particular regimes of physical interest.
Such regimes correspond to different asymptotic regions \cite{JMP14} in the half plane 
$(A,B) \in \mR\times[0,\infty)$, namely:
\begin{enumerate}
\item
the asymptotic region $A^2+B^2 \to \infty$ with $A < -B$ (i.e.\ $(A,B)$ below the ``critical line'' $A+B=0$), 
corresponds to a regime of high temperatures, where the Fermi-Dirac distribution is well approximated by a 
Maxwell-Boltzmann distribution;
\item
the asymptotic region $A^2+B^2 \to \infty$ with $A > -B$ corresponds to the limit $T\to0$, in which case
we speak of  ``degenerate fermion gas'';
\item
the asymptotic region $A^2+B^2 \to \infty$ with $A \sim B$ (i.e.\ $(A,B)$ approaches the critical line $A+B=0$)
corresponds to a ``collimation regime'', $\abs{\uu} \to 1$,  where the velocities of the electrons are all
aligned along a ($(x, t)$-dependent) direction in the $\pp$-space (the direction
determined by $\uu$); there are two types of collimation, depending on whether the critical line is approached from
below (Maxwell-Boltzmann collimation) or from above (degenerate gas collimation); 
\item
opposite to the collimation limit, the asymptotic region $B \to 0$ corresponds to the diffusive limit $\abs{\uu} \to 0$, where the velocities
are randomly spread over all directions.
\end{enumerate}
\par
The asymptotic analysis of equations \eqref{MEgraph} in these regimes is based on the following result, 
which  has been proven in Ref.\ \cite{JMP14}.
\par
\smallskip
\noindent
{\bf Theorem.}
The functions $\cI_N^s$ have the following asymptotic behavior:
\begin{enumerate}
\item
in the Maxwell-Boltzmann limit, $A^2+B^2 \to \infty$, with $A < -B$,
\BE
\label{BFasym1}
  \cI_N^s(A,B) \sim  \e^A\,I_N(B),
\EE
where $I_N$ are the modified Bessel functions of the first kind;
\item
in the degenerate gas limit, $A^2+B^2 \to \infty$, with $A > -B$,
\BE
\label{BFasym2}
\cI_N^s(A,B) \sim
  \frac{1}{\pi\Gamma(s+1)} \int_0^{\mathrm{C}(A,B)} \hspace{-16pt} \cos(N\theta) (A+B\cos\theta)^s d\theta,
\EE
where
\BE
\label{Cdef}
  \mathrm{C}(A,B)
   = \Re\left[\cos^{-1}\!\left(-\textstyle{\frac{A}{B}}\right)\right]
   = \left\{
  \begin{aligned}
  &\arccos\left(-\textstyle{\frac{A}{B}}\right),& &\text{if $-B < A < B$},
  \\
  &\pi,& &\text{if $A \geq B$}.
  \end{aligned}
  \right.
  \EE
\end{enumerate}
\subsection{Maxwell-Boltzmann regime}
The Maxwell-Boltzmann regime is the limit for large $T$ 
and corresponds to $A^2+B^2 \to \infty$ with $A < -B$ in the $(A,B)$ half plane.
%Since, in this case, $C\abs{\pp} \gg \nu(\pp)\cdot\BB + A$, then 
%the MEP-state \eqref{MEPstate} is well approximated by the Maxwellian-like distribution 
%\BE
%\label{MEPstateMB}
%  f^\eq = \exp\left[-C\abs{\pp} + \nnu(\pp) \cdot\BB  + A\right],
%\EE
%and the approximation \eqref{BFasym2} holds.
In this case, we can use the approximation \eqref{BFasym1}.
Note, in particular, that in such limit the functions $\BF_N^s$ become factorized and independent 
on the index $s$.
Then, the constraint equations \eqref{CE2} become
\BE
\label{CE3}
\begin{aligned}
  &\e^A I_0(B)n_T = n, 
\\[4pt]
  &\frac{I_1(B)}{I_0(B)} = \abs{\uu},
\\[4pt] 
  &2k_BT = e.
\end{aligned}
\EE
and it can be shown that the the MEP-state \eqref{MEPstate} is well approximated  by the
Maxwellian-like distribution
\BE
\label{MEPstateMB2}
  f^\eq = \frac{n}{n_T \,I_0(B)} \exp\left[-\frac{c}{k_BT}\abs{\pp} + B\,\nnu(\pp) \cdot \frac{\uu}{\abs{\uu}}\right],
\EE
where
\BE
\label{Bu}
B = \Big(\frac{I_1}{I_0}\Big)^{-1}(\abs{\uu}).
\EE
Moreover, we get the explicit form of $P_{ij}$, $Q_{ij}$ and $S_j$:
\BE
\label{PQSMB}
     \begin{aligned}
     &P_{ij} = \frac{n}{\abs{\uu}^2}\left[ X(\abs{\uu})\, u_i u_j  
     + \left(1-X(\abs{\uu})\right) u_i^\perp u_j^\perp \right],
     \\[6pt]
     &Q_{ij} = \frac{2c}{e}\frac{n}{\abs{\uu}^2}\left[ X(\abs{\uu})\, u_i^\perp u_j^\perp  
     + \left(1-X(\abs{\uu})\right) u_i u_j \right],
     \\[6pt]
     &S_j = neu_j
      \end{aligned}
\EE
where
$$
  X(\abs{\uu}) = \frac{I_0(B) + I_2(B)}{2I_0(B)}
$$
and $B$ is given by \eqref{Bu}.
\par
By playing a little with the asymptotic expansions of the modified Bessel functions $I_n$ we obtain the 
asymptotic behavior of $X(\abs{\uu})$ in the diffusive limit:
\BE
\label{asymX0}
   X(\abs{\uu}) = \frac{1}{2} + \frac{1}{4}\abs{\uu}^2 + \cO(\abs{\uu}^4),
   \qquad 
   \text{as $\abs{\uu} \to 0$,}
\EE
and in the collimation limit:
\BE
\label{asymX1}
   X(\abs{\uu}) = 1 - 2(1-\abs{\uu})^2 + \cO\big((1-\abs{\uu})^3\big),
   \qquad 
   \text{as $\abs{\uu} \to 1$.}
\EE
\par
Substituting $S_j = neu_j$ in the third of the moment equations \eqref{MEgraph} yields, after 
a little algebra,
\BE
\label{eMB}
  \pt_te + cu_j\pt_j e \pm c u_j \pt_jV  = 0.
\EE
Thus, the isothermal case ($e = 2k_BT$ constant) is only compatible with $u_j \pt_jV  = 0$,
i.e.\ the component of the force field parallel to the velocity field must vanish.
In this case, the pseudo-momentum balance equation reduces to
\begin{multline}
\label{MEgraph2}
\pt_t (n u_i) + c\pt_j\left(\frac{nX(\abs{\uu})\, u_i u_j}{\abs{\uu}^2}  
     + \frac{n\left(1-X(\abs{\uu})\right) u_i^\perp u_j^\perp}{\abs{\uu}^2} \right)
\\[6pt]
 \pm   \frac{c}{k_BT} \frac{n}{\abs{\uu}^2} X(\abs{\uu})\, u_i^\perp u_j^\perp \pt_j V = 0.
\end{multline}
\subsection{Degenerate gas regime} 
The degenerate gas regime is the limit for $T\to 0$ and corresponds to $A^2+B^2 \to \infty$ with $A > -B$,
in the $(A,B)$ half plane.
In this case, we can use the approximation \eqref{BFasym2}--\eqref{Cdef}.
It is convenient to put
$$
   A = R\cos\psi, \qquad B = R\sin\psi,
$$
and rewrite \eqref{BFasym2} as follows:
\BE
\label{T0apprxPolar}
\cI_N^s(R\cos\psi,R\sin\psi) \sim R^s \cF_N^s(\psi), \qquad R > 0, \quad 0 \leq  \psi  < \frac{3\pi}{4},
\EE
where
\BE
\label{cFdef}
\cF_N^s(\psi) =  \frac{1}{\pi\Gamma(s+1)} \int_0^{\mathrm{C}(\psi)} \hspace{-16pt} \cos(N\theta) (\cos\psi+\sin\psi\,\cos\theta)^s d\theta
\EE
and
\BE
\label{Cdef2}
 \mathrm{C}(\psi)
    = \Re\left[\cos^{-1}\!\left(-\cot\psi\right)\right]
   = \left\{
  \begin{aligned}
  &\arccos\left(-\cot\psi \right),& &\text{if $\frac{\pi}{4} < \psi < \frac{3\pi}{4}$},
  \\[2pt]
  &\pi,& &\text{if $0\leq \psi \leq \frac{\pi}{4}$}.
  \end{aligned}
  \right.
\EE
The asymptotic form of the constraint equations \eqref{CE2} is now 
\BE
\label{ABT0}
\begin{aligned}
  &\cF_0^2(\psi) R^2 n_T = n, 
  \\[3pt]
  &\frac{\cF_1^2(\psi)}{\cF_0^2(\psi)} = \abs{\uu},
  \\[3pt]
  &\frac{\cF_0^3(\psi)}{\cF_0^2(\psi)} k_BTR = e.
\end{aligned}
\EE
Note that:
\begin{enumerate}
\item
$\abs{\uu}$ only depends on $\psi$ and we can write
\BE
\label{psifromu}
  \psi = \Big(  \frac{\cF_1^2}{\cF_0^2}  \Big)^{-1}(\abs{\uu});
\EE
\item
recalling \eqref{nTdef}, from the first of the above equations we have $R \sim 1/T$ and,
then, the third equation shows that $e$ remains positive even though $T\to0$.
\end{enumerate}
From the above considerations  it is readily seen that, in the limit $T\to 0$, the MEP-state \eqref{MEPstate} takes
the typical degenerate Fermi-Dirac form
\BE
\label{MEPstateDG}
      f^\eq = \theta\left[ -\sqrt{\frac{2\pi \hbar^2 n} {\cF_0^2(\psi)}} \,\abs{\pp} + \nu(\pp)\cdot \frac{\uu}{\abs{\uu}}\sin\psi 
    + \cos\psi    \right],
\EE
where $\theta$ denotes the Heaviside function and $\psi(\uu)$ is given by \eqref{psifromu}.
Using  \eqref{PQexpr}, \eqref{T0apprxPolar} and \eqref{ABT0} we obtain
the following expressions of $P_{ij}$ and $Q_{ij}$ and $S_j$ for a degenerate electron gas:
\BE
\label{PQSDG}
\begin{aligned}
&P_{ij} = \frac{n}{\abs{\uu}^2} \left[ Y(\abs{\uu})u_iu_j + \left(1-Y(\abs{\uu})\right)u_i^\perp u_j^\perp \right],
\\[4pt]
&Q_{ij} =  \frac{\sqrt{n}}{\hbar\sqrt{\pi}\abs{\uu}^2} 
\left[ Z(\abs{\uu}) u_iu_j + Z_\perp(\abs{\uu}) u_i^\perp u_j^\perp \right], 
\\[4pt]
&S_j =  W(\abs{\uu}) \,\frac{neu_j}{\abs{\uu}},
\end{aligned}
\EE
where
$$
\begin{aligned}
  &Y(\abs{\uu}) = \frac{\cF_0^2(\psi) + \cF_2^2(\psi) }{2\cF_0^2(\psi)},
  &\quad
  &Z(\abs{\uu}) = \frac{\cF_0^1(\psi) - \cF_2^1(\psi) }{2\sqrt{2\cF_0^2(\psi)}},
\\
  &Z_\perp(\abs{\uu}) = \frac{\cF_0^1(\psi) + \cF_2^1(\psi) }{2\sqrt{2\cF_0^2(\psi)}},
  &\quad
  & W(\abs{\uu}) = \frac{\cF_1^3(\psi)}{\cF_0^3(\psi)} ,
\end{aligned}
$$
and $\psi = \psi(\abs{\uu})$ is given by Eq.\ \eqref{psifromu}.
\par
By using the techniques developed in Ref.\ \cite{JMP14}, it is not difficult to calculate the  asymptotic
behavior of the functions $Y(\abs{\uu})$, $Z(\abs{\uu})$, $Z_\perp(\abs{\uu})$ and $W(\abs{\uu})$ in the
two limits $\abs{\uu} \to 0$ (diffusion) and $\abs{\uu} \to 1$ (collimation).
\par
For $\abs{\uu} \to 0$ we obtain:
\BE
\label{YZu0}
\begin{aligned}
   Y(\abs{\uu})& = \frac{1}{2} + \frac{1}{8}\abs{\uu}^2 + \cO(\abs{\uu}^4),
\\[4pt]
  Z(\abs{\uu})& =  \frac{1}{2} - \frac{1}{8}\abs{\uu}^2+ \cO(\abs{\uu}^4),
\\[4pt]  
    Z_\perp(\abs{\uu})& = \frac{1}{2} - \frac{1}{8}\abs{\uu}^2+ \cO(\abs{\uu}^4),
\\[4pt] 
  W(\abs{\uu})& = \frac{3}{2}\abs{\uu} + \cO(\abs{\uu}^3).
\end{aligned}
\EE
\par
For $\abs{\uu} \to 1$ we obtain
\BE
\label{YZu1}
 \begin{aligned}
   Y(\abs{\uu})& = 1 - 2(1-\abs{\uu}) + \cO\big((1-\abs{\uu})^2\big),
   \\[4pt]
  Z(\abs{\uu})&  = \frac{(14)^\frac{5}{4}}{\sqrt{30\pi}}\,(1 - \abs{\uu})^\frac{5}{4}
  + \cO\big((1-\abs{\uu})^\frac{9}{4}\big),
    \\[4pt]
    Z_\perp(\abs{\uu})& = \frac{\sqrt{5}\, (14)^\frac{1}{4}}{\sqrt{6\pi}}\,(1 - \abs{\uu})^\frac{1}{4}
  + \cO\big((1-\abs{\uu})^\frac{5}{4}\big),
  \\[4pt] 
  W(\abs{\uu})& = 1 - \frac{7}{9}\,(1 - \abs{\uu})
  + \cO\big((1-\abs{\uu})^2\big).
  \end{aligned}
\EE
\subsection{Collimation regime}
The collimation limit corresponds to the absence of spread in  the particle directions, i.e.\ to $\abs{\uu} \to 1$.
It can be shown \cite{JMP14,MorandiBarletti14} that this limit is equivalent to $A^2+B^2 \to \infty$ with $A/B \to -1$.
However, there is a completely different behavior when the critical line $A = - B$ is approached from below (Maxwell-Boltzmann
collimation) of from above (degenerate gas collimation).
\par
\smallskip
The first case corresponds to taking the limit $B\to \infty$ in the ``Maxwellian'' distribution  \eqref{MEPstateMB2},
which produces a delta in the angle between $\pp$ and $\uu$, namely
\BE
\label{MEPstateMB2}
  f^\eq = \frac{2\pi\, n}{n_T} \exp\left[-\frac{c}{k_BT}\abs{\pp}\right] \delta\left(\nu(\pp) - \frac{\uu}{\abs{\uu}} \right).
\EE
Moreover, since $X(\abs{\uu}) \to 1$ as $\abs{\uu} \to 1$ (see Eq.\ \eqref{asymX1}), 
from Eq.\ \eqref{PQSMB} we obtain
$$
  P_{ij} \to nu_iu_j, \qquad Q_{ij} \to \frac{2c}{e} n u_i^\perp u_j^\perp
$$
and the pseudo-momentum balance equation reduces to
$$
  \pt_t (n u_i) + c\pt_j (nu_iu_j)   \pm \frac{2c}{e}\, nu_i^\perp u_j^\perp  \pt_jV = 0.
$$
By using the continuity equation $\pt_t n + c\pt_j(nu_j)$ the latter can  be rewritten as
\BE
\label{MBPE}
   \pt_t u_i + cu_j \pt_j u_i  \pm  \frac{2c}{e} u_i ^\perp u_j^\perp   \pt_jV = 0
\EE 
which is decoupled from the continuity equation for $n$.
As pointed out in Refs.\ \cite{JMP14,MorandiBarletti14},
this equation reveals that collimated electrons in graphene have the properties of a geometrical-optics system,
with ``refractive index'' 
$$
  N(\xx) = \e^{\mp\frac{2}{e} V(\xx)} =  \e^{\mp\frac{1}{k_BT} V(\xx)}.
$$
By also considering the energy balance equation \eqref{eMB}, we finally obtain the system
\BE
\left\{
 \begin{aligned}
 &\pt_t u_i + cu_j \pt_j u_i  \pm  \frac{2c}{e} u_i ^\perp u_j^\perp   \pt_jV = 0,
\\[4pt]
 &\pt_te + cu_j\pt_j e \pm c u_j \pt_jV  = 0.
 \end{aligned}
  \right.
\EE
\par
\smallskip
In order to derive the collimation equations for a degenerate gas, we start from the expression \eqref{PQSDG} of $P_{ij}$, $Q_{ij}$ and $S_j$,
and use the asymptotic relations \eqref{YZu1} to obtain that
$$
  P_{ij} \to nu_iu_j, \qquad Q_{ij} \to 0, \qquad S_j \to neu_j,
$$
as $\abs{\uu} \to 1$.
But then, the hydrodynamic system \eqref{MEgraph} degenerates into the decoupled system
$$
\left\{
        \begin{aligned}
        &\pt_t n  + c\pt_j (n u_j) = 0,
        \\[4pt]
        &\pt_t u_i + cu_j \pt_j u_i  = 0,
        \\[4pt]
        &\pt_te + cu_j\pt_j e \pm c u_j \pt_jV  = 0.
        \end{aligned}
        \right.
$$
Such a ``trivial'' asymptotic behavior of collimated degenerate electrons has already been pointed out in Ref.\ \cite{JMP14} in the
isothermal case, and is due to the vanishing effective-mass tensor $Q$.
\subsection{Diffusion regime}
The diffusion regime corresponds to the limit $\abs{\uu}\to 0$ of vanishing mean velocity.
In order to observe the diffusive behavior we have to introduce in \eqref{MEgraph} 
a current-relaxation term $-n\uu/\tau$, and to rescale time and velocity as 
$$
  t^* = \tau t, \qquad \uu^* =  \frac{1}{\tau}\uu.
$$
In this way we obtain the system
\BE
\label{MEscal2}
\left\{
\begin{aligned}
&\pt_{t^*} n  + c\pt_j (n u_j^*) = 0,
\\[3pt]
&\tau^2 \,\pt_{t^*} (n u_i^*) + c\pt_j P_{ij}  \pm Q_{ij} \pt_j V = - nu_i^*
\\[3pt]
&\tau \pt_{t^*} (n e) + c\pt_j S_j \pm c \tau nu_j^* \pt_jV  = 0,
\end{aligned}
\right.
\EE
where the terms $P_{ij}$, $Q_{ij}$ and $S_j$, depending only on $\uu/\abs{\uu} = \uu^*/\abs{\uu^*}$, remain unchanged
except that the Lagrange multipliers must satisfy
\BE
\label{ABdiff}
  \frac{\cI_1^2(A,B)}{ \cI_0^2(A,B)} = \tau\abs{\uu^*}.
\EE
In the diffusive limit $\tau \to 0$ we obtain the condition $\cI_1^2(A,B) = 0$, which is satisfied if and only if $B = 0$ \cite{JMP14}.
Since 
\BE
  \BF_N^s(A,0) = 
  \left\{
  \begin{aligned}
  &\phi_s(A),&\quad &\text{if $N = 0$,}
  \\
  &0,&\quad &\text{if $N \geq 1$,}
  \end{aligned}
  \right.
\EE
from the first of \eqref{CE2} with $B=0$ we obtain 
\BE
\label{invA}
  A = \phi_2^{-1}\left( \frac{n}{n_T} \right)
\EE
and, moreover,
\BE
\label{PQSdiff}
\begin{aligned}
  &P_{ij}(A,0) = \frac{n}{2}\, \delta_{ij},
\\[3pt]
  &Q_{ij}(A,0) = \frac{c\, n_T}{2k_BT}\, \phi_1\Big( \phi_2^{-1}\Big( \frac{n}{n_T} \Big) \Big)\, \delta_{ij},
\\[3pt]
  &S_j(A,0) = 0.
 \end{aligned}
\EE
Letting $\tau\to0$ in Eq.\ \eqref{MEscal2} yields, therefore, the diffusive system
\BE
\label{Diff1}
\left\{
\begin{aligned}
&\pt_{t^*} n  + c\pt_j (n u_j^*) = 0,
\\[3pt]
& nu_i^* = -  \left( c\pt_j P_{ij}  \pm  Q_{ij} \pt_j V \right),
\end{aligned}
\right.
\EE
with $P = P(A,0)$ and $Q = Q(A,0)$ given by \eqref{PQSdiff}, that is, in terms of the original time variable
\BE
\label{Diff2}
\pt_t n = \frac{\tau_0 c^2}{2} \pt_j \left[ \pt_j n 
   \pm \frac{n_T}{k_BT}\,\phi_1\Big( \phi_2^{-1}\Big( \frac{n}{n_T} \Big) \Big)\pt_j V \right].
\EE
\smallskip
It is not difficult to check that the diffusion equation \eqref{Diff2} take the specific form
 \BE
\label{Diff3}
 \pt_t n = \frac{\tau_0 c^2}{2} \pt_j 
 \left(\pt_j n \pm \frac{n}{k_BT}\,\pt_j V \right)
\EE
in the Maxwell-Boltzmann limit and 
\BE
\label{Diff4}
 \pt_t n = \frac{\tau_0 c}{2}\, \pt_j 
 \left(c\pt_j n \pm \frac{1}{\hbar\,\sqrt{\pi}}\,\sqrt{n}\,\pt_j V \right).
\EE
in the degenerate gas limit.
\par
\smallskip
We remark that, owing to the conical dispersion relation  \eqref{conic}, 
the drift-diffusion equations \eqref{Diff2},  \eqref{Diff3} and  \eqref{Diff4} have a ``specular'' structure with respect to the drift-diffusion equations for Fermions with the usual parabolic dispersion relation 
\cite{TrovatoReggiani10,JungelKrausePietra11,JSP12}.
Indeed, the diffusion coefficient (which is proportional to the variance of the velocity distribution), is here
independent of the temperature $T$, because the particles move with constant speed $c$, while it is proportional 
to $T$ in the parabolic case.
On the other hand, the mobility coefficient (which is related to the distribution of the second derivative 
of the energy, i.e.\ to the effective-mass tensor) is here temperature-dependent 
while in the parabolic case is constant. 
Also the nonlinearity, which in the parabolic case affects the diffusive term, in Eqs.\ \eqref{Diff1} and 
\eqref{Diff4} is found in the drift term. 
\section{Conclusions}
\label{S6}
We have presented the systematic derivation from the Maximum Entropy Principle of hydrodynamic equations 
describing a population of electrons subject to spin-orbit interactions. 
In the second part of the paper we have treated more extensively the case of electrons on a single-layer graphene sheet.
\par
The hydrodynamic equations have the form of a Euler-like system of conservation laws for density, $n$, momentum, $\uu$, and energy, $e$,
in each of the two bands (the band indices are here omitted).
Such system is of hyperbolic character, which ensures its (at least) local well-posedness.
It is worth to remark that the full nonlinear structure of the MEP-state is retained, so that no assumptions of linear response or 
quasi-isotropic distribution are needed.  
\par
The system, in general, is not explicitly closed, i.e.\ no explicit constitutive relations, expressing the higher-order moments 
$P_{ij}$, $Q_{ij}$ and $S_j$ as functions of $n$, $\uu$ and $e$, can be given.
However, in the case of graphene and for particular asymptotic regimes (namely, the limits of high and zero temperature, the limit 
of collimated direction and the diffusive limit), the closure is fully explicit.
\par
As already mentioned in the Introduction, our results are not able to capture the physics of the system when the semiclassical
approximation is not valid, that is when the quantum coherence becomes important.
This typically happens in presence of rapid potential variations, such as potential steps or barriers.
In these cases one expects the equations derived here to be a good approximation in a ``semiclassical region'',  
far enough from the potential steps (constituting instead the ``quantum region'').
The semiclassical regions could be coupled to the quantum ones by means of quantum-classical interface conditions, analogous
to those developed for standard, i.e.\ scalar and parabolic, particles (see Ref.\ \cite{DEA02} and references therein).

\section*{Acknowledgements}
 This work has been partially supported by INdAM-GNFM, Progetto Giovani Ricercatori 2013 
{\em Quantum fluid dynamics of identical particles:  analytical and numerical study}.

\end{document}